\documentclass{aa}  
\usepackage{graphicx}
\usepackage{txfonts}
\usepackage{hyperref}
\usepackage{multirow}

\usepackage{xcolor}
\newif\ifrevision
\revisionfalse   % <-- clean version (uncomment for final "clean" version)

\ifrevision
  \newcommand{\rev}[1]{\textcolor{blue}{\textbf{#1}}}
  \newcommand{\edrev}[1]{\textcolor{orange}{#1}}             % editorial revision
  \newcommand{\langrev}[1]{\textcolor{red}{#1}}             % language revision
\else
  \newcommand{\rev}[1]{#1}
  \newcommand{\edrev}[1]{#1}
  \newcommand{\langrev}[1]{#1}
\fi

\newcommand{\del}[1]{\ifrevision\textcolor{gray}{\sout{#1}}\fi}
\usepackage[normalem]{ulem}

\begin{document} 
    
    \titlerunning{Slow focus sensor for the Keck I  LGS-AO system using focal plane wavefront sensing}
    \authorrunning{Salgueiro, R. et al.}

   \title{Slow focus sensor for the Keck I laser guide star adaptive optics system using focal plane wavefront sensing}

   \subtitle{}

   \author{Rafael M. Salgueiro
          \inst{1,2},
          Carlos M. Correia
          \inst{3,4},
          Benoit Neichel
          \inst{1},
          Antonin Bouchez
          \inst{5},
          Peter Wizinowich
          \inst{5},
          Avinash Surendran 
          \inst{5},
          Max Service
          \inst{5},
          Thierry Fusco 
          \inst{1,6},
          Cédric Taïssir 
          \inst{1,7},
          Pierre Jouve
          \inst{1,8}
          }

   \institute{Aix-Marseille Univ, CNRS, CNES, LAM, 13013 Marseille, France 
        \and
             Macquarie University, Balaclava Rd, Macquarie Park NSW 2113, Australia
         \and
             Faculdade de Engenharia da Universidade do Porto, Rua Dr. Roberto Frias s/n, 4200-465 Porto, Portugal
        \and
        Center for Astrophysics and Gravitation, Instituto Superior Técnico, Av. Rovisco Pais 1, 1049-001 Lisboa, Portugal
        \and 
             W.M. Keck Observatory, 65-1120 Mamalahoa Hwy, Waimea, HI 96743, USA
        \and 
             DOTA, ONERA, Université Paris Saclay, 91120 Palaiseau, France
         \and 
             DOTA, ONERA, 13330, Salon-de-Provence, France
        \and 
         Space ODT, Av. da França 492, 4050-277 Porto, Portugal
             }

   \date{Received September 29, 2025; accepted\del{November 16, 2025} \edrev{February 02, 2026}}
 
  \abstract
   {Laser guide stars (LGSs) have been deployed for the last 20-30 years in ground-based astronomical telescopes to overcome the limited sky coverage of classical adaptive optics (AO) systems. Unfortunately, slow altitude drifts of the sodium layer compromise focus measurements, generating the so-called slow focus error, and, consequently, a natural guide star (NGS) is needed to compensate that error. The Keck I telescope AO system uses a 20x20 Shack-Hartmann (SH) wavefront sensor (WFS) for slow focus tracking (with a 5x5 mode used on fainter stars). This approach is far from optimal due to limited sky coverage, since the available NGSs are usually very faint.}
  % aims
   {Our goal is to develop a different technique for slow focus tracking and make it fully operational using focal plane wavefront sensing (FPWFS), which can significantly increase sky coverage and allow slow focus tracking at higher frequencies, reducing the lag error. The Keck I near-infrared (NIR) tip-tilt sensor, known as TRICK, is used to obtain the focal plane images without any hardware modifications being necessary.}
  % methods
   {We develop, characterize, and compare three different FPWFS algorithms, namely Gerchberg-Saxton (GS), linearized focal plane technique (LiFT), and Gaussian fit (Gf). These algorithms are studied for the specific purpose of slow focus sensing in the NIR (H and K bands) using numerical simulations and data collected at Keck in 2025 (bench and on-sky).}
  % results
   {The three algorithms were studied and characterized against different criteria such as linearity, computational costs, and resistance to low signal-to-noise ratio and/or residuals. From the results obtained, the main candidate for an on-sky deployment was GS, for which the main deciding factor was its higher stability and robustness under the presence of residuals. For that reason, on-sky tests in closed loop were made with GS.}
  % conclusions
   {On-sky tests showed promising results, with GS successfully compensating for purposely introduced focus errors, even under the presence of high turbulence conditions. These tests represent an important step toward the full operationalization of this tool, expected in the coming months. This work can also be extrapolated to other existing 8-10 m class telescopes, or even future 30-40 m class telescopes, where the use of FPWFS can significantly improve sky coverage and reduce the lag error.}


   \maketitle
%
%-------------------------------------------------------------------

\section{Introduction}

Adaptive optics (AO) systems have been deployed in recent decades in ground-based astronomical telescopes to increase the angular resolution of astronomical images, which otherwise would be highly degraded by atmospheric turbulence (\cite{roddier1999adaptive}; \cite{hardy1977real}; \cite{rousset1990first}). In classical natural guide star (NGS) AO systems, the NGS must be within $\sim$ 30" of the science target for adequate correction in the near-infrared (NIR) \citep{fried1982anisoplanatism}. However, the availability of sufficiently bright NGSs is reduced, which limits NGS AO usage to a few percent of the sky (\cite{plantet2022sky}; \cite{fusco2006sky}; \cite{le1998sky}).

To overcome the limited sky coverage, laser guide stars (LGSs) were proposed in 1982 for ground-based telescopes \citep{happer1994atmospheric}. These artificial stars are created using lasers tuned to 589.2 nm to excite the sodium atoms present in the sodium layer ($\sim$ 90 km of altitude), which when returned to their ground state emit the desired \edrev{back-scattered} radiation. Although a much higher sky coverage can be achieved using LGSs, an NGS is still needed; however, it can be fainter, since it is only required for tip-tilt (TT) \citep{rigaut1992laser}, and it can be further off-axis due to the larger isokinetic angle \citep{hardy1998_book}. This TT star can also be used to perform slow focus tracking, an error that originates from altitude drifts of the sodium layer altitude (\cite{hickson2006focus}; \cite{pfrommer2010high}).

The slow focus tracking will be addressed in this article; namely, for a real telescope application at the Keck I telescope in Hawaii. Currently, in the Keck I LGS-AO system \citep{wizinowich2006wm}, the slow focus tracking is done using a Shack-Hartmann (SH) wavefront sensor (WFS) with 20x20 \edrev{sub-apertures} (or its 5x5 mode), which in terms of sky coverage is not the best approach since high sky coverage requires usage of faint NGSs. The proposed alternative is to use focal plane wavefront sensing, which allows one to increase sky coverage, and the frequency at which we track slow focus, also reducing the lag error.

The idea at Keck I is to use the images from the NIR TT sensor, i.e., TRICK (\cite{wizinowich2014near}), not only for TT sensing, but also for slow focus tracking using focal plane wavefront sensing techniques to retrieve focus. Three different algorithms were studied to this end; namely, Gerchberg-Saxton (GS), the linearized focal plane technique (LiFT), and \edrev{Gaussian} fit (Gf).

In the first part of this paper, the algorithms are described in section 2, and a description of the Keck I AO system is presented in section 3. The results from the different studies performed toward a novel slow focus sensor at Keck I are divided into numerical simulations (section 4), \langrev{daytime bench tests} (section 5), and on-sky tests (section 6). In section 7, the conclusions and future prospects are given.
   
 \section{Focal plane wavefront sensing}
 
Focal plane wavefront sensing (FPWFS) techniques are inverse methods that allow one to estimate the aberrated phase from an image at the focal plane. The image irradiance ($I_i$), given an object irradiance ($I_0$), considering incoherent radiation and monochromatic light is given by the following equation:
  
\begin{equation}\label{Eq_img_model_with_noise}
I_i = p \ast I_0 + n\, ,
\end{equation}

\noindent where $\ast$ represents the convolution operation, $n$ is the noise, and $p$ (known as the point spread function) is given by

\begin{equation}\label{Eq_psf}
p = \left| \mathcal{F}\{U \, e^{i(\Phi+\Phi_d)}\} \right|^2 \, .
\end{equation}

\noindent $U \, e^{i(\Phi+\Phi_d)}$ is the electric field at the pupil plane, where $\Phi$ is the phase we want to estimate, and $\Phi_d$ the phase diversity. The phase diversity allows one to break the sign ambiguity for even modes (assuming a symmetric pupil), such as focus \citep{gonsalves1982phase}. 

\begin{figure}[h]
\centering
\includegraphics[scale=0.52]{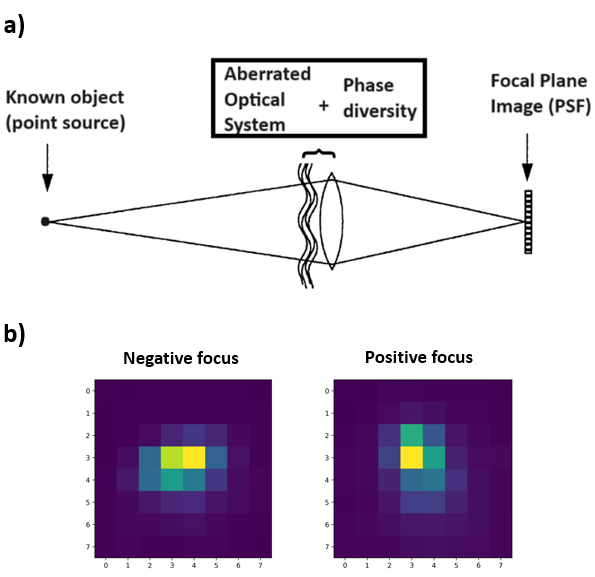}
\caption{a) Phase retrieval scheme (Adapted from \cite{roddier1999adaptive}). b) Example from the NIR TT sensor at Keck I (TRICK), illustrating how the presence of astigmatism (0º) as phase diversity causes the focal plane image to stretch in perpendicular directions depending on the focus sign, thereby resolving the focus sign ambiguity.}
\label{phase_retrieval_scheme}
\end{figure}

There are different categories of focal plane techniques, as is illustrated in Tab. \ref{table:paxmann_classification}, where the Paxmann classification is used. In this article, we only work on phase retrieval algorithms, for which the object is known (point source - star), and the system aberrations are unknown as depicted in Fig. \ref{phase_retrieval_scheme} a). Some FPWFS algorithms are also referred to in the literature as phase diversity or joint estimation, but in this article phase diversity always refers to an amount of phase introduced to break the even modes degeneracy and not to FPWFS methods.

\begin{table}[ht]
\centering
\caption{\edrev{Paxman classification of inverse problems.}}
\label{table:paxmann_classification}
\begin{tabular}{|c|c|c|}
\hline
\multirow{2}{*}{\textbf{System}} & \multicolumn{2}{c|}{\textbf{Object}} \\
\cline{2-3}
 & \textbf{Known} & \textbf{Unknown} \\
\hline
\textbf{Known} & Omniscience & Image restoration \\
\hline
\textbf{Unknown} & \begin{tabular}[c]{@{}c@{}}Phase retrieval\\ (System identification)\end{tabular} & 
\begin{tabular}[c]{@{}c@{}}Joint estimation\end{tabular} \\
\hline
\end{tabular}
\tablefoot{\edrev{Following the classification proposed by \citet{paxman2007}.}}
\end{table}

The main advantage of FPWFS techniques (as presented in Fig. \ref{phase_retrieval_scheme}) is that they are full-pupil techniques, which is particularly important when dealing with very faint stars as usually happens with TT stars in LGS-AO systems. In addition, for a given magnitude, FPWFS can also allow one to perform phase measurements at a higher rate, since we can achieve a high signal-to-noise ratio (S/N) much faster compared to using conventional methods as the SH WFS. These advantages make the FPWFS an excellent solution for slow focus tracking in LGS-AO systems. Also, no additional hardware is required for the Keck case, because of existing diversity (i.e., astigmatism) in the beam, which is very convenient (see Fig. \ref{phase_retrieval_scheme} b)).

However, there are challenges associated with FPWFS, the main one being the nonlinearity between the phase, $\Phi$, and the focal plane image, $I_i$, as is shown in Eq. \ref{Eq_img_model_with_noise} and \ref{Eq_psf}. This implies that nonlinear optimization algorithms need to be used. Nonlinear optimization algorithms search for a solution (set of parameters) that minimizes a particular metric error. We start from an initial guess for the solution and then update it at each iteration, usually using gradient-based (first order) search algorithms. Local minima are a common problem encountered in nonlinear optimization, and the starting point (first guess) can play a very important role in mitigating this problem \citep{bjorck2024numerical}.

Conventionally, our first phase guess for the FPWFS methods would be the phase diversity itself. However, non-common path aberrations (NCPAs) are usually present at a significant level, and taking them into account for the first guess can significantly improve the final solution, as has recently been explored by Arseniy Kuznetsov et al. \citep{arseniy2024}. To that end, in-focus images are used to determine the NCPAs, to be taken into account as the first guess (plus the phase diversity). This method can be seen as a phase diversity calibration step.

Some other aspects that affect the FPWFS performance are sub-Nyquist sampling, low S/N, and the presence of high-order residuals. In the case of slow focus tracking for LGS-AO systems, these challenging conditions are usually the ones we have to deal with \citep{van2006wm}. Finally, the computational costs associated with nonlinear techniques can sometimes be an issue. \langrev{However, despite these challenges, we show that FPWFS can still be used in such challenging conditions for the specific purpose of slow focus tracking, with promising results for the Keck case study.}

\del{Despite these challenges, we show that FPWFS can still be used in such challenging conditions for the specific purpose of slow focus tracking, with promising results for the Keck case study.}
 
\section{Phase retrieval methods}

In this section, we describe the three algorithms investigated for slow focus tracking at Keck; namely, GS, LiFT, and Gf. All three are single-image phase-retrieval algorithms, meaning that the phase estimate is obtained from a single focal-plane image. The TRICK images, which include astigmatism, enable the breaking of the even-mode degeneracy, in particular the focus mode (see Fig. \ref{phase_retrieval_scheme}).

The initial phase estimate was obtained using an in-focus image, whereby each algorithm was applied to obtain a phase estimate containing the initial phase diversity (astigmatism) plus any other NCPAs present. This diversity calibration \citep{arseniy2024} was performed offline and the resulting calibrated phase diversity was used as the initial guess for all subsequent algorithm runs.

\subsection{Gerchberg-Saxton (GS)}

The GS algorithm is the most standard and classical phase retrieval algorithm \citep{gerchberg1972}. In GS, we iterate back and forth between the pupil plane and the focal plane, knowing that the electric field at the focal plane is the Fourier transform of the electric field in the pupil plane. To make the inverse path, we \langrev{use} the inverse Fourier transform. A priori constraints in each plane \langrev{are} also applied at each iteration. In the pupil plane, the constraint involves applying the pupil mask, and at the focal plane, enforcing the amplitude value. The amplitude value at the focal plane is equal to $\sqrt{p}$ assuming a point source object and ignoring noise components (see Eq. \ref{Eq_psf}), and since we measure $p$, we can use that constraint.

The initial guess for GS is the phase diversity (plus any other NCPAs present), and after iterating we converge on the final solution. Although at first glance GS does not directly apply any nonlinear algorithm, Fienup showed that GS is closely related to the steepest descent method \citep{fienup1982}, and also proved that the Frobernius norm of the difference between the estimated focal plane intensity ($u^2$), and the measured one ($u_0^2 = \sqrt{p}$) \langrev{either decreases or remains constant} at each iteration, n. \langrev{This norm is defined as:}

\begin{equation}
    E_n^2 =  || u_0^2 - u^2 ||_F \, .
\end{equation}

\noindent \del {can only stay the same or decrease at each iteration.} \rev{For a more detailed description of the specific GS algorithm used, please see Appendix \ref{Appendix_GS_algo}}.

\subsection{Linearized focal plane technique (LiFT)}

The linearized focal plane technique (LiFT) is a more recent FPWFS method \citep{meimon2010lift}, for low-order mode sensing such as TT and focus in the context of LGS-AO systems. LiFT allows one to estimate a phase that \edrev{minimizes} the difference between the recorded image, $I_i$, and the model image (obtained using Eq. \ref{Eq_img_model_with_noise} and \ref{Eq_psf}) using a maximum likelihood estimation assuming Gaussian statistics for the noise. 

The LiFT algorithm is discussed elsewhere in detail (see \cite{meimon2010lift}; \cite{plantet2014phd}; \cite{plantet2013experimental}; \cite{arseniy2024}). Overall, LiFT can be seen as a standard nonlinear algorithm, since it seeks to minimize a certain metric error, using a gradient search algorithm iteratively. LiFT uses a small phase approximation (\cite{gonsalves2001small}), which is similar to using a gradient search algorithm.

\subsection{Gaussian fit (Gf)}

Since the goal of this paper is to use FPWFS to perform slow focus tracking, a more simplistic algorithm was also explored to that end, i.e., Gf. A point source image with astigmatism (0°) as phase diversity, which will be the case for the Keck study, will stretch in perpendicular directions according to the sign of the input focus (see Fig. \ref{phase_retrieval_scheme} b)). By performing a \edrev{Gaussian} fit to the focal plane image, we can use the obtained values for the standard deviations ($\sigma_x$ and $\sigma_y$) to estimate the focus value ($Z_4$) by using the following expression:

\begin{equation}\label{Eq_Gf}
    Z_4 =  sign(astig.) \left( \frac{\sigma_y}{\sigma_x} - 1 \right) \times \alpha \, ,
\end{equation}

\noindent where $\alpha$ is a gain that needs to be calibrated in simulations or on the AO bench. We show that the linear relationship between $Z_4$ and $\sigma_y/\sigma_x$ is valid either in numerical simulations or with real data.

Contrary to GS and LiFT, where the image model was given by Eq. \ref{Eq_img_model_with_noise} and \ref{Eq_psf}, for Gf the image model is simply a 2D Gaussian. However, a nonlinear least square error was still used to perform the \del{GF} \langrev{Gf} \citep{scipy_least_squares}, with the error metric being the difference between the recorded image and the modeled image. Similar fitting algorithms have been explored (e.g. \cite{tokovinin2006donut}) for low-order mode sensing; namely, focus.

\section{Keck study case}
 
In this article, we study three alternatives to the low-bandwidth WFS (LBWFS), a 20x20 SH that is currently used for slow focus tracking in the Keck I LGS-AO system. These alternatives are the FPWFS methods presented in the previous section. The ultimate goal is to use the NIR TT sensor, i.e., TRICK, to perform slow focus tracking with one of these methods, which can significantly improve the sky coverage and reduce the lag error. Some previous attempts to implement this idea with LiFT were made in the past \citep{plantet2016lift}.

Quantitatively, transitioning from a 20×20 SH to a FPWFS corresponds to an increase of $\sim$ 300 times in the collected flux (F) for a given exposure time comparing with a single SH sub-aperture, taking into account all SH sub-apertures within Keck pupil. This corresponds to a magnitude gain of roughly $-6.2$ as given by $\Delta m = -2.5 \, \log_{10} (F/F_{\mathrm{ref}})$ \citep{pogson1856magnitudes}. In addition, we note that the LBWFS only receives 10 \% of the visible light, the other 90 \% going to the STRAP visible TT sensor (not used here), meaning that the flux and magnitude gains are even higher, i.e., $\sim$ 3000, and $\Delta m = -8.7$, respectively. Of course, this is just a rough estimate, since we neglect the relative flux of the star in the two bands (visible and NIR) or the higher S/N required for FPWFS compared to calculating a centroid. Nevertheless, it provides an indication of the potential flux and magnitude gains achievable with FPWFS for slow focus tracking in the Keck I LGS-AO system, relative to the current LBWFS approach.

The angular pixel size for the TRICK camera is 50 mas, which corresponds to a sampling lower than 0.5 Nyquist both for K and H bands (allowed bands in TRICK) as shown in table \ref{table:nyquist}. The point source images obtained by TRICK are also “naturally” astigmatic, aberration being introduced by transmission through a dichroic, which can be perfectly used as phase diversity to remove the focus sign ambiguity. The amount of astigmatism  present ($Z_6$ Zernike polynomial \citep{noll1976zernike}, also known as astigmatism 0°) is $\sim$ 200 nm rms \citep{wizinowich2014near}.

\begin{table}[h]
\caption{\edrev{Angular pixel sizes corresponding to different samplings.}} 
\label{table:nyquist} 
\centering
\begin{tabular}{c c c c} 
\hline\hline
        & 1 Nyquist & 0.5 Nyquist & 0.25 Nyquist \\ 
\hline
H-band & 16      & 31        & 62         \\
K-band & 20      & 41        & 82         \\
\hline
\end{tabular}
\tablefoot{\edrev{Angular pixel sizes in mas corresponding to 1.0, 0.5, and 0.25 sampling for the H and K bands. The Keck I telescope diameter is 10.949 m, $\lambda(H) = 1.65 \, \mu m$, and $\lambda(K) = 2.18 \, \mu m$.}}
\end{table}

\begin{figure*}[ht]
\centering
\includegraphics[scale=0.45]{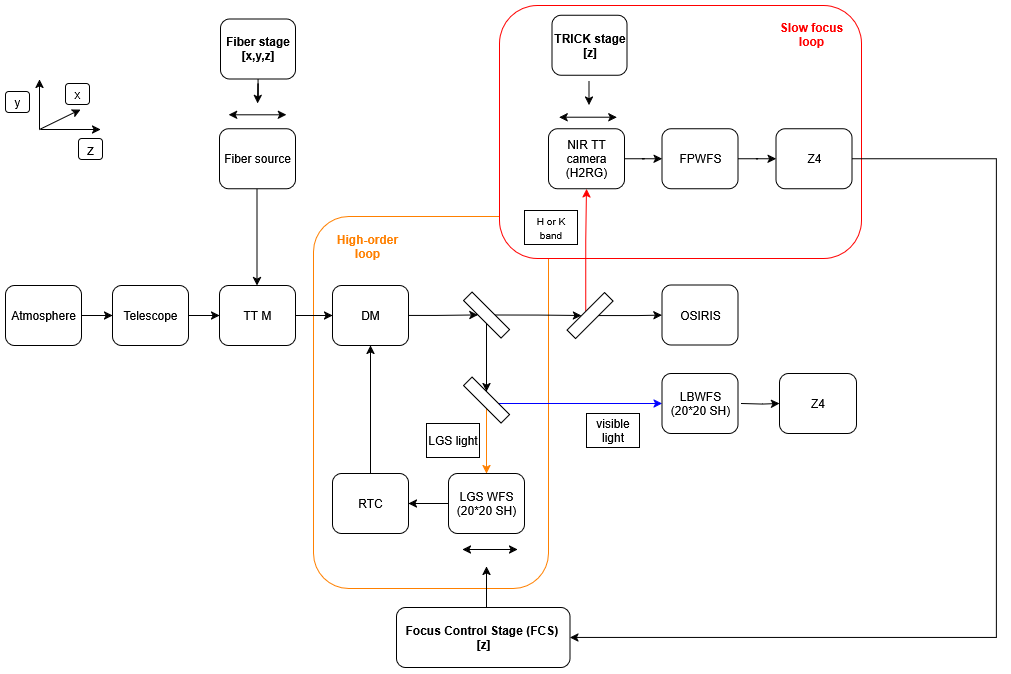}
\caption{Keck I LGS-AO system scheme (\cite{wizinowich2006wm}, \cite{van2006wm}. The high-order (all modes expect TT) loop is operated with the light coming from the LGSs, while the low-order loop is operated with NGS light. In the low-order loop, the TT and slow focus ($Z_4$) sensing can be both performed using images provided by TRICK. The current approach for slow focus sensing is to use the LBWFS. The measured focus values are then used to compensate for the focus error by moving the FCS. OSIRIS is the science camera.}
\label{keck_ao_scheme}
\end{figure*}

In Fig. \ref{keck_ao_scheme}, the Keck I AO scheme is presented with its main components (\cite{wizinowich2006wm}; \cite{van2006wm}). The two main represented parts of the AO system can be divided in the high-order loop from the LGS WFS (SH WFS) to the deformable mirror (DM) via the real time computer (RTC), and the slow focus loop from the TT sensor (TRICK) to the WFS focus stage (FCS). Three focus stages are available in the AO system: the fiber stage precision control unit (PCU), the LGS WFS focus stage (FCS), and the TRICK focus stage. These focus stages were used in the bench tests to introduce certain amounts of focus.

The defocus distance ($d$) is related to the root mean square (rms) defocus coefficient ($Z_4$) by the following expression \citep{blanc2003calibration}:

\begin{equation}\label{Eq_z4_conversion}
    Z_4 (rad) = \frac{\pi d}{8 \sqrt{3} \lambda (F/D)^2} \, ,
\end{equation}

\noindent where $\lambda$ is the wavelength, $F$ is the focal length, and $D$ is the telescope diameter. In the Keck case, we have $D = 10.949$ m and $F = 150$ m.

The slow focus tracking using TRICK is expected to be performed under challenging conditions, namely sub-0.5 Nyquist sampling, low S/N since the NGS is usually very faint, and high-order residuals that increase as the angular separation between the NGS and LGS increases. However, these are the expected conditions of real LGS-AO systems, making \del{it} \langrev{TRICK configuration} a perfect scenario to test FPWFS for slow focus tracking. These results can then be extrapolated to other 8-10 m class telescopes or even future extremely large telescopes (ELTs).

In addition to sampling, noise and residuals, other important aspects should be taken into account to access performance, including the algorithms' linear range, computational costs, and performance with small field-of-view (FoV) images. The small field is due to the need to reduce readout noise via multiple readouts of the TRICK detector (\cite{castella2016status}, \cite{wizinowich2014near}).

\section{Numerical simulations}

In this section, numerical simulations tests characterizing GS, LiFT, and Gf performance according to different criteria are presented. The simulations were done in a Keck-like environment (telescope diameter, TRICK sampling of 50 mas, 200 nm rms of astigmatism (0°) as diversity, etc.) as described in the previous section.

The first aspect to be considered when doing slow focus tracking is to know how fast we should correct this focus error, which originates from altitude drifts from the sodium layer located at $\sim$ 90 km of altitude. Different studies were conducted to characterize the temporal variations of the sodium layer altitude, and the power spectral density (PSD) of the mean altitude variations is well fit by the following model (\cite{pfrommer2010high}; \cite{neichel2013characterization}):

\begin{equation}\label{Eq_PSD_delta_h}
\text{PSD}(f) = \alpha f^{\beta} \, .
\end{equation}

\noindent These altitude variations introduce a focus error ($\sigma_{Na}$) in the system given by \citep{pfrommer2010high}

\begin{equation}\label{Eq_sigma_Na}
    \sigma_{Na} = \frac{1}{16\sqrt{3}} \left( \frac{D}{h-h_0} \right)^2 \Delta h \, ,
\end{equation}

\noindent with $h$ being the mean altitude of the sodium layer (90 km), $h_0$ the telescope altitude (4145 m), and $D$ the telescope diameter (10.949 m). It is important to note that $\sigma_{Na}$ is proportional to $D^2$, meaning that slow focus tracking needs to be performed significantly faster for 30-40 m telescopes to achieve the same performance as in 8-10 m class telescopes.

\begin{figure}[h]
\centering
\includegraphics[scale=0.50]{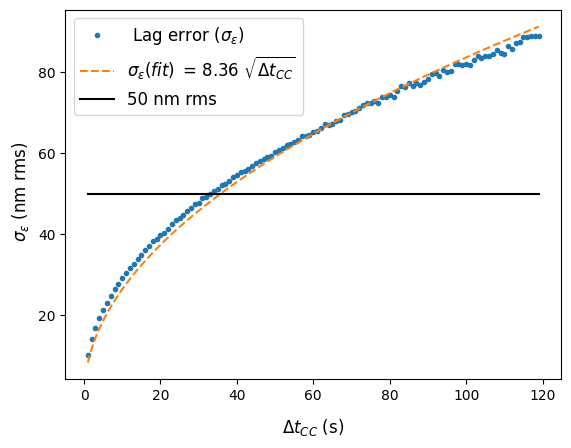}
\caption{Lag error ($\sigma_{\epsilon}$) according to the correction cadence ($\Delta t_{CC}$). Perfect correction is assumed, i.e., only the \del{lag}temporal error is taken into account. The values of $\sigma_{\epsilon}$ are proportional to $\sqrt{\Delta t_{CC}}$.}
\label{lag_error}
\end{figure}

Using a value of $\alpha = 35 m^2Hz^{-1}$, and $\beta=-1.9$, values found experimentally by \citep{neichel2013characterization}, a 5 hour simulated altitude variation series was calculated from the PSD in Eq. \ref{Eq_PSD_delta_h}, and the corresponding slow focus series from Eq. \ref{Eq_sigma_Na}. Then, for different correction cadence times ($\Delta t_{CC}$), the original defocus series $\sigma_{Na}$ was averaged over non-overlapping intervals of duration $\Delta t_{CC}$, and the resulting block-averaged estimate was shifted by $\Delta t_{CC}$ to simulate the slow focus controller response. The difference between the lag-and-averaged focus series and the original one gives the rms lag focus error ($\sigma_{\epsilon}$), with the results being presented in Fig. \ref{lag_error}. From the results obtained, a focus correction every 35 s gives an rms lag error of 50 nm, which can be reduced to $\sim$ 18 nm rms if we perform the focus correction every 5 s. Reducing the lag error is one of the possibilities of FPWFS, since it is a full-pupil technique and high S/N can be achieved faster than with conventional WFS such as the SH WFS.

\begin{figure}[h]
\centering
\includegraphics[scale=0.50]{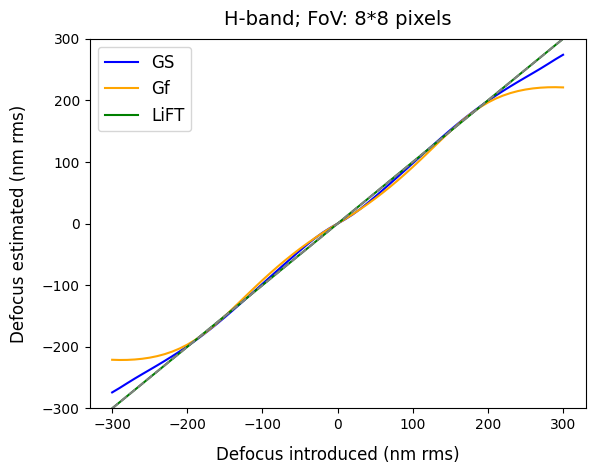}
\caption{Focus ramp simulation under ideal conditions (no noise or high-order residuals). Results presented for H band, with the angular pixel size of the simulated images being 50 mas.}
\label{focus_ramp__fov_8_ideal_simulation}
\end{figure}

Another important aspect when characterizing the FPWFS algorithms is their linear range. From the results given in Fig. \ref{lag_error}, when performing performing corrections every 35 s, we expect a $\sigma_{\epsilon}$ of 50 nm rms, which would ideally require the FPWFS methods to have a linear range of (-150, +150) nm rms, considering a 3$\sigma$ interval. The FPWFS methods are expected to perform slow focus tracking much faster, so the linearity requirements would be lower after correcting for any initial large focus error when the loop is first closed. To study the algorithms' linear range, a focus ramp simulation was performed, with the results shown in Fig. \ref{focus_ramp__fov_8_ideal_simulation}. These results show that all algorithms are linear at least within $\pm$ 200 nm rms.

It should be noted that the GS algorithm is being applied with a sampling error model, i.e., the given images have an angular pixel size of 50 mas (below 0.5 Nyquist), but a model of 1.0 Nyquist sampling is used (standard GS algorithm is suited for at least 1.0 Nyquist sampled images). In GS algorithm, we iterate back and forth between the pupil and focal planes, and a zero padding factor of 2 is used for the pupil when using a 1 Nyquist model, while the focal plane image, which is sampled below 0.5 Nyquist, is zero-padded so that it has the same dimensions as the zero-padded pupil \rev{(see Appendix \ref{Appendix_GS_algo} for a more detailed description)}. Although at first this was not expected to work since a considerable model error was being made, simulation results showed that the linear behavior in the focus ramps is well maintained. Of course, a gain needs to be applied (4.75 gain in Fig. \ref{focus_ramp__fov_8_ideal_simulation} for GS) to compensate for this sampling error. This gain shows to be stable and robust, since similar gains were also obtained in focus ramps in bench tests, as is shown in the next section. For the moment, a theoretical explanation for this behavior has not yet been found, it being for now an empirical result.

Another interesting aspect is that when introducing a similar sampling error in the LiFT algorithm, the linearity is also maintained and with a similar gain value, showing that both LiFT and GS can estimate focus even when \rev{committing} considerable sampling errors. To further explore this behavior, we studied the corresponding gains to be applied in focus ramps according to the sampling error committed, with the results shown in Appendix \ref{Appendix_sampling_error}. We emphasize that all LiFT results presented in the following sections were obtained using the correct sampling in the LiFT model, i.e., an angular pixel size of 50 mas. The inclusion of sampling errors in LiFT is considered only in Appendix \ref{Appendix_sampling_error}.

\begin{figure}[h]
\centering
\includegraphics[scale=0.50]{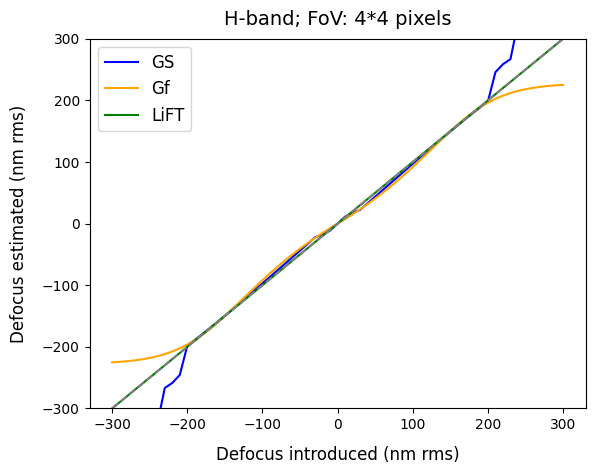}
\caption{Focus ramp simulated under the same conditions as the one presented in Fig. \ref{focus_ramp__fov_8_ideal_simulation}, but with a smaller FoV (4x4 pix.).}
\label{focus_ramp__fov_4_ideal_simulation}
\end{figure}

A similar study to the one presented in Fig. \ref{focus_ramp__fov_8_ideal_simulation} was made, but with \del{low FoV}\langrev{small-FoV} images, given that one of the TRICK modes reads only 4x4 pixels around the center of gravity of the image. As we can see \rev{in Fig. \ref{focus_ramp__fov_4_ideal_simulation}}, the linear behavior is still very satisfactory within $\pm$ 200 nm rms, and the applied optimal gains for GS and Gf are very similar to the ones obtained before. This shows that all algorithms can be used even with \del{low FoV}\langrev{small-FoV} images.

\begin{figure}[h]
\centering
\includegraphics[scale=0.50]{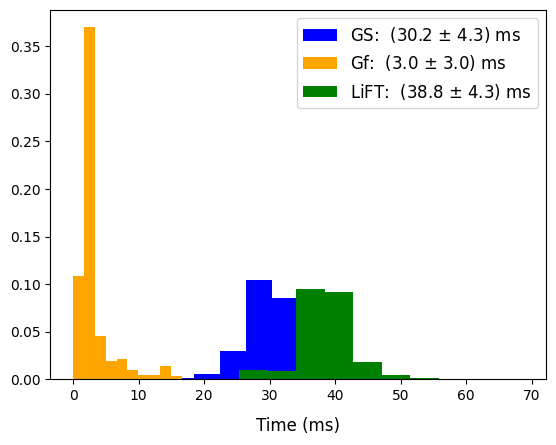}
\caption{Histogram of time elapsed per focus estimate for GS, Gf, and LiFT on a personal computer. A total of 20 iterations was made both for GS and LiFT. For Gf, which uses a scipy method to perform the Gf, we only iterated until convergence.}
\label{hist_time_elapsed}
\end{figure}

Computational costs are also an important factor when accessing the algorithms' performances, and a simulation of the time spent per focus estimation was made, with the results presented in Fig. \ref{hist_time_elapsed}. In this simulation, 1000 samples were used, with focus values introduced from a Gaussian distribution ($\mu = 0$, $\sigma = 50$ nm rms), and under ideal conditions (no noise or high-order residuals). In LiFT, and GS, which perform Fourier transforms from pupil plane to focal plane, a pupil with 16x16 pixels was found to be the optimal trade-off between speed and focus estimation accuracy. In fact, the focus estimation is only affected when going down to a 8x8 pupil size, where aliasing effects highly degrade the algorithms' performances.

Other than pupil sampling, for LiFT, the number of estimated modes also affects the computational costs, i.e., the more modes we want to estimate, the more computational costs. For GS, that does not happen since we estimate a phase rather than modes. In order to make a fair comparison, the computational costs of LiFT were studied for LiFT estimating four modes (TT, focus, and astigmatism (45°)), with the results being shown in Fig. \ref{hist_time_elapsed}. \del{The results shown in Fig. \ref{hist_time_elapsed} are for LiFT estimating four modes (TT, focus, and astigmatism (45°)).} Still, LiFT is intended for low-order mode sensing, so these times per focus estimate are \del{on}\langrev{within} the expected order of magnitude. \langrev{For the slow focus tracking case, which is the application envisioned in this paper for FPWFS, all algorithms allow several focus estimates per second, meaning that slow focus tracking could be done every 1 s or even faster without any problem provided we have sufficient S/N to perform focus estimation.}

\begin{figure}[h]
\centering
\includegraphics[scale=0.50]{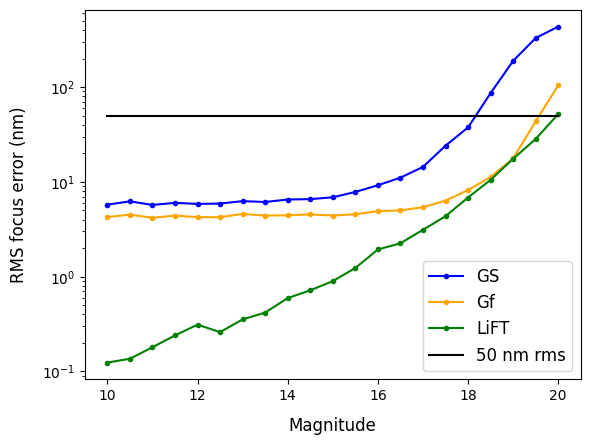}
\caption{Limiting magnitude (H-band) simulations considering photon and readout (RON) noise. No high-order residuals are considered here. LiFT does not commit a sampling error and for that reason, the focus errors are almost irrelevant for high-S/N cases. For GS and Gf, that does not happen since they do not have a “perfect” model.}
\label{sky_coverage}
\end{figure}

Finally, we examined the limiting magnitude for each algorithm, as the ability to use fainter stars directly increases sky coverage, the main goal of employing FPWFS for slow focus tracking. The results are shown in Fig. \ref{sky_coverage}. For each magnitude value, 200 noisy images were generated with random focus values (generated from a Gaussian distribution ($\mu = 0$ and $\sigma = 50$ nm rms)), which were then compared with the focus values retrieved from GS, Gf, and LiFT. According to the given magnitude, the short-exposure (SE) time given was adapted, ranging from 1 ms for $m=$ 10-12 up to 10 ms for $m>16$. The SE noisy frames were then numerically integrated with long-exposure (LE) times ranging from 0.5 s for $m<16$ and 1.0 s for $m>16$. This is similar to what happens in the Keck operational environment, where TRICK SE times are adapted according to the NGS magnitude, and numerical LE times can be adapted to obtain a higher S/N.

From the results obtained, we can see that for a threshold error of 50 nm rms, we have the following H-band limiting magnitudes: $m=18$ for GS, $m=19.5$ for Gf, and $m=20$ for LiFT. The LBWFS can be used up to $m=19$ by making use of its 5x5 mode, but at the cost of having very long integration times (up to 120 s), which affect the focus measurement made and also greatly increases the lag error. Taking an intermediate example, for a magnitude between 15.5 and 16, an integration time of 30 s is used for the LBWFS. If, on the other side, a FPWFS was used, a much lower integration time could be used; for instance, 1 s. This would represent a reduction in the lag error of $\sim$ 37 nm rms. This lag error reduction increases with magnitude.

\section{Bench tests at Keck}

Previous tests toward using the TRICK sensor for slow focus tracking were made in the past 10 years at Keck; namely, with the LiFT algorithm \citep{plantet2016lift}. Most of the previous datasets (focus ramps) coming from these tests were \edrev{analyzed} as part of a master thesis project \citep{salgueiro2024master}, with the three presented FPWFS algorithms: GS, Gf, and LiFT. The results were promising, and novel tests toward operationalization of a FPWFS method for slow focus tracking were planned for 2025. Next, the results from these tests will be presented.

\subsection{Focus ramps}

The first study made on the Keck AO bench was focus ramps in open loop to verify \del{that}the linear behavior, and the gains of the algorithms for a first “sanity” check. The focus was introduced using the fiber stage (PCU) with the low- and high-order loops both open, and the conversion to focal distance was made using Eq. \ref{Eq_z4_conversion}. To obtain the in-focus position on TRICK a full width at half-maximum (FWHM) fit was performed, and the approximate position giving the lowest value was considered as the zero-focus position. 

The images taken in the in-focus position were used to perform the diversity calibration step \citep{arseniy2024} to take NCPAs present into account. This phase estimate with in-focus images was used as a first guess for all other images, and highly improved the algorithms' performances. This had already been verified with previous datasets. In the case of Gf, the correct term would be bias calibration, since no phase estimate is done in Gf, but for the sake of clarity we also refer to it as a diversity calibration.

\begin{figure}[h]
\centering
\includegraphics[scale=0.50]{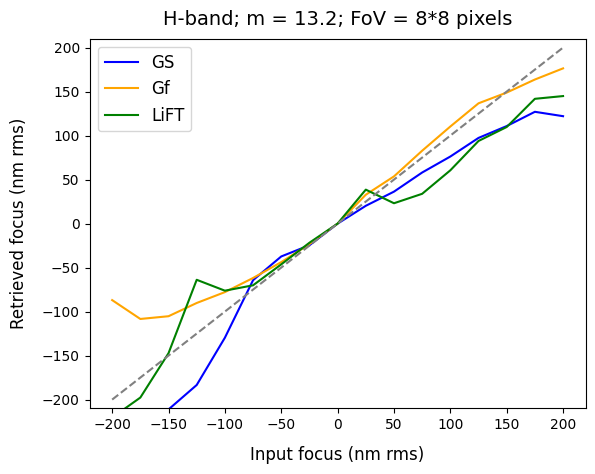}
\caption{Focus ramp example for the H band.}
\label{focus_ramp_bench_H_band}
\end{figure}

In Fig. \ref{focus_ramp_bench_H_band}, we show an example of a bench focus ramp in the H band. As qA expected, the linearity is lower than numerically predicted in simulation, where no noise or high-order residuals were present. Gf showed higher linearity in the bench focus ramps overall. Based on this focus ramp, the LiFT and GS linearity is still within 50 nm rms, which when performing slow focus tracking every 1-5 s can be perfectly sufficient.

The gains verified for GS and Gf in the focus ramps are very similar to what was obtained from previous datasets, which shows consistency. For GS, the gain approximately varies between 4.5 and 5.5, and for Gf the gain is $\sim$ 600 for the K band and $\sim$ 350 for the H band (see Eq. \ref{Eq_Gf}).

Finally, we note that the focus ramps results obtained for the K band show larger linearity ranges, and a better performance overall. This is explained by the fact that the images in the K band are better sampled. Still, the H band is usually the band used with TRICK, since K-band light usually goes to the science camera (OSIRIS), and for that reason we focus on H-band results in this paper.

\subsection{Slow focus in open loop}

As a next step forward, slow focus tests in open loop were made. To that end, a numerically generated slow focus series of 5 minutes with 1 second time steps was generated using Eq. \ref{Eq_PSD_delta_h}, \ref{Eq_sigma_Na}. The slow focus was introduced using the LGS WFS focus control stage (FCS) with both low- and high-order loops closed. Note that the low-order loop was closed in terms of TT, but the slow focus loop was open. Every second, a 1 s TRICK image was saved, and then the results were processed to see how well the algorithms followed the curve.

\begin{figure}[h]
\centering
\includegraphics[scale=0.50]{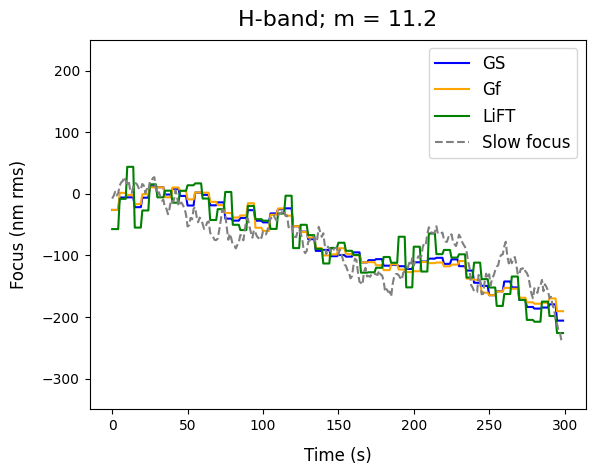}
\caption{Slow focus results, without AO residuals, after bias and gain calibration. Results obtained with 5 numerically integrated 1-second TRICK frames. Linear fit parameters from raw results: GS ($m=0.75$, $b=24.7$ nm, $R^2=0.79$); Gf ($m=0.28$, $b=11.4$ nm, $R^2=0.78$); LiFT ($m=0.71$, $b=18.0$ nm, $R^2=0.71$). The given linear fit parameters are the slope (m), the bias (b), and the regression fit ($R^2$).}
\label{slow_focus_no_residuals}
\end{figure}

\begin{figure}[h]
\centering
\includegraphics[scale=0.38]{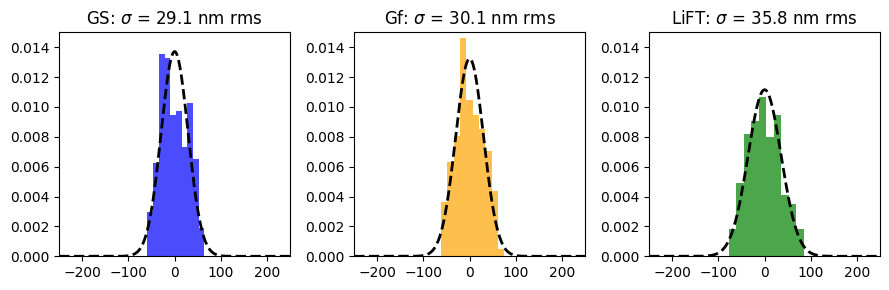}
\caption{Histogram of focus error for each algorithm from the results presented in Fig. \ref{slow_focus_no_residuals}.}
\label{slow_focus_no_residuals_hist_error}
\end{figure}

In Fig. \ref{slow_focus_no_residuals}, we have an example without introducing high-order residuals. The results shown are the ones after performing a bias and gain calibration. This bias and gain calibration was necessary since the diversity calibration was performed with the first image of the dataset, which was clearly not in focus as it should be. The gain and bias values for the linear fit from the raw data are listed in the Fig. \ref{slow_focus_no_residuals} caption. An interesting aspect is that, although there is a bias and gain issue, the algorithms' linearity is very well maintained, as is seen from the $R^2$ values. The histogram error for each algorithm (after calibration) is presented in Fig. \ref{slow_focus_no_residuals_hist_error}, where we can see that all algorithms show a similar performance, i.e., an rms error of around 30 nm.

\begin{figure}[h]
\centering
\includegraphics[scale=0.50]{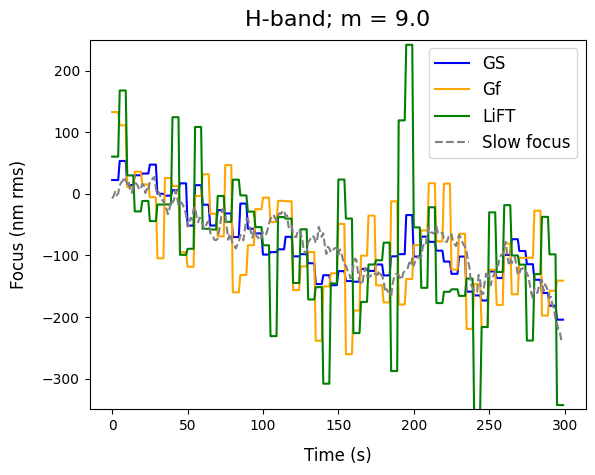}
\caption{Slow focus results, with AO residuals, after bias and gain calibration. Results obtained with five numerically integrated 1-second TRICK frames. Linear fit parameters from raw results: GS ($m=0.23$, $b=-6.2$ nm, $R^2=0.74$); Gf ($m=0.05$, $b=-6.2$ nm, $R^2=0.40$); LiFT ($m=0.22$, $b=-29.9$ nm, $R^2=0.22$). The given linear fit parameters are the slope (m), the bias (b), and the regression fit ($R^2$).}
\label{slow_focus_with_residuals}
\end{figure}

\begin{figure}[h]
\centering
\includegraphics[scale=0.38]{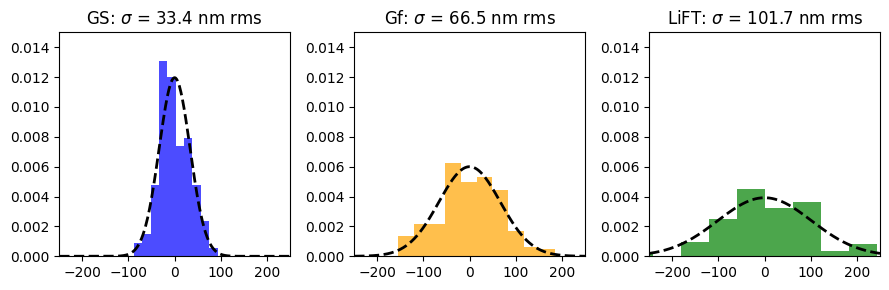}
\caption{Histogram of focus error for each algorithm from the results presented in Fig. \ref{slow_focus_with_residuals}.}
\label{slow_focus_with_residuals_hist_error}
\end{figure}

A slow focus test in open loop considering the presence of high-order residuals ($\sim$ 200 to 300 nm rms) was also made. \rev{These residuals were introduced by reading a time series of DM commands that simulate time-varying turbulence and closely reproducing median turbulence conditions at the Keck site.} This case is much closer to what is expected in real scenario conditions on-sky, and the obtained results are presented in Fig. \ref{slow_focus_with_residuals}. Once more, the presented results already account for the bias and gain calibration, as was explained for the previous case. The performance decline is clearly visible, especially for Gf and LiFT, which show much more instability. As we can see in Fig. \ref{slow_focus_with_residuals_hist_error}, GS clearly has the lowest rms error, with a value of 33 nm rms, while Gf and LiFT have rms errors of 66 and 102 nm, respectively. Another big difference in terms of linearity can be seen in the $R^2$ values (see Fig. \ref{slow_focus_with_residuals} caption), where GS has a value of 0.74, while Gf and LiFT have 0.40 and 0.22, respectively. High-order residuals decrease the linearity of LiFT and Gf much more than for GS. These results show that the GS algorithm is more robust to the presence of high-order residuals and also more stable, which is a very important advantage when considering an on-sky implementation.

The presence of high-order residuals also indicates a change in the algorithms’ gain (see the $m$ values from the linear fits reported in the captions of Figs. \ref{slow_focus_no_residuals} and \ref{slow_focus_with_residuals}). However, for an operational closed-loop scenario, the most critical aspect is the linearity of the response, which in the GS case remains well preserved, as was discussed in the previous paragraph. Although the gain is modified, this effect is mitigated in closed-loop operation, particularly in a slow focus loop where the focus evolves slowly.

Recent work from \cite{arseniy2024} at the Very Large Telescope (VLT) demonstrated an on-sky focus ramp with very good focus estimation and linearity. Yet, some key differences exist in relation to the Keck TRICK case. The VLT sampling is more than twice Nyquist ($\sim$ 4.4 pixels per $\frac{\lambda}{D}$) at H-band versus under 0.5 Nyquist for TRICK. Also, the diversity calibration method used by Kuznetsov et al. was not LiFT itself, which may have improved linearity of the focus ramp. Although LiFT does not exhibit the best performance compared to GS in the TRICK configuration, LiFT should not be excluded as a viable slow focus sensor under different operational conditions.

Another aspect in which LiFT exhibits a slight limitation concerns the number of modes to be estimated. Bench tests indicate that estimating a larger number of modes (e.g., the first 8–10) does not necessarily improve performance and, in some cases, may even degrade it due to model over-fitting and cross-talk between modes, which tends to be more prominent with under-sampled images as happens with TRICK. In an operational scenario, determining the optimal number of modes can therefore be challenging.

Based on all the results obtained from numerical simulations and bench tests, Tab. \ref{table:performance} was compiled to compare the algorithms. Gf showed higher linear ranges overall, especially in bench focus ramps. However, since the slow focus tracking is expected to be done fast (every 1-5 s) a linear range of $\pm$ 50 nm rms can be more than enough for the algorithms, in which case all algorithms are almost equivalent. All three algorithms are capable of performing with small-FoV images, but Gf is more stable in such conditions. All algorithms are capable of performing several focus estimates per second, so this is not a limiting factor for any of the algorithms. Under low-S/N conditions, LiFT and Gf showed a better performance overall, even in numerical simulations. Finally, the performance under high-order residuals is the aspect where the algorithms' performances sets them apart. GS showed a considerably better performance under the presence of high-order residuals by a significant factor, as was shown in the slow focus tests. The incorporation of the high-order residuals PSD into LiFT or GS was not investigated in this work. Given the strong performance of GS in the presence of high-order residuals, such incorporation was not deemed necessary at this stage. Nevertheless, it may represent an interesting avenue for future investigation, although care would be required, as the associated regularization could \del{influence/bias}\langrev{influence and/or bias} the final solution.

Overall, analyzing all performance parameters, GS is shown to be the main candidate for an on-sky operationalization for slow focus tracking with the Keck I LGS-AO system. For that reason, on-sky tests were performed with GS in an attempt to fully demonstrate its capabilities in a real operational environment. Results are shown in the next section.

\begin{table}
\caption{Algorithms' performances from numerical simulations and bench tests.}  
\label{table:performance}    
\centering                          
\begin{tabular}{c c c c}        
\hline\hline                 
 & GS & Gf & LiFT \\    
\hline                        
   Linear range & \checkmark & \checkmark \checkmark & \checkmark \\      
   Low FoV & \checkmark & \checkmark \checkmark     & \checkmark  \\
   Computational costs & \checkmark & \checkmark     &  \checkmark \\
   Noise & \checkmark & \checkmark \checkmark  & \checkmark \checkmark \\
   Residuals & \checkmark \checkmark & x    & x \\
\hline                                   
\end{tabular}
\end{table}

\section{On-sky tests at Keck}\label{on_sky_section}

The slow focus tracking closed-loop results with the Keck I LGS-AO system are discussed in this section. These tests were made on the night of February 8, 2025, in the last hour of the half night (5 A.M. to 6 A.M. local time), with \langrev{an} H-band, bright TT \del{stars}\langrev{star} ($m_V = 10$), and under challenging seeing conditions ($\sim 1 ''$ in the K band). On the day before, calibration images (in the H band) were collected and were then used for the diversity calibration step. The focus corrections were sent to the FCS focus stage (high-order WFS focus stage) to compensate for the slow focus error in real time. The poor seeing made it hard to keep the star on TRICK that night, which contributed to only $\sim$ 10 min. of on-sky data. The TRICK images SE time was 1 ms, and by numerical integration 1 s exposure images were then used as inputs for the GS algorithm, meaning the slow focus loop frequency was 1 Hz.

\begin{figure}[h]
\centering
\includegraphics[scale=0.50]{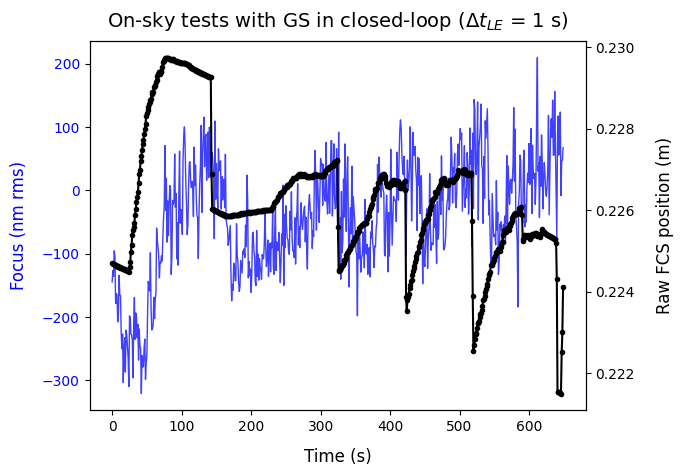}
\caption{On-sky results from the night of February 8, 2025, with the GS algorithm for slow focus tracking in closed loop. Raw FCS positions are also shown, in which the four moments when an intentional focus was introduced can be clearly identified. The mean and standard deviation of focus values obtained by GS neglecting the first 100 s, when the loop was still converging, were, respectively, -22 and 71 nm rms. The negative mean was expected, since during the experiment the out-of-focus movements were always made in the same direction (negative focus).}
\label{on_sky_fcs_plus_GS}
\end{figure}

The GS focus measurements obtained from the on-sky tests are shown in Fig. \ref{on_sky_fcs_plus_GS}. During the first 100 seconds, we can clearly see that the GS corrected focus was slowly converging on zero, meaning that the slow focus loop was correctly compensating for the initial focus error verified before closing the loop. Then we intentionally put the system out of focus by moving the FCS stage, as is seen by the FCS positions in Fig. \ref{on_sky_fcs_plus_GS}, to test the algorithm's recovery of focus. We can see that after those movements, GS successfully identified the focus error and compensated for it since the FCS position starts returning approximately to its initial state. We point out that the FCS “in-focus” position is expected to slowly change with the telescope elevation as the target is tracked, which explains why after the out-of-focus movements we do not return to exactly the same position. It should also be noted that the slow focus loop gain was varied during the experiment, being initially low and then being slowly increased as the close loop stability was verified. For this reason, the focus compensation speed varied throughout the experiment, gradually increasing as the loop gain was raised.

These results also showed the incredible stability of GS algorithm, which even under very bad seeing conditions did not diverge, i.e., the retrieved values being mainly within $\pm$ 100 nm rms. This confirms the predicted GS robustness to high-order residuals, which was seen during the slow focus bench tests with residuals.

\section{Conclusions}

In this paper, we have explored the possibility of using FPWFS for slow focus tracking in the LGS-AO Keck I system. Three different algorithms were explored to this end, namely GS, LiFT, and Gf, with a comparison study being made through numerical simulations and several bench tests. The main performance criteria where the algorithms differ by a considerable factor is the robustness under the presence of high-order residuals. This was verified in the slow focus bench tests with AO residuals where the regression fit for GS was $R^2 = 0.74$, while for Gf and LiFT was 0.40, and 0.22 respectively. These results indicate that GS is more robust in those conditions, with a much lower linearity decline and focus error overall.

The on-sky tests in closed loop performed with GS showed promising results toward operationalization of the tool. GS correctly compensated for different amounts of intentionally introduced focus even under the presence of challenging seeing conditions ($\sim$ 1'' in the K band), confirming its robustness and stability under the presence of high-order residuals. Further on-sky tests are planned to fully assess the robustness and performance of this technique under different conditions, including a direct comparison between the LBWFS and GS algorithms. If GS is shown to outperform the LBWFS, a novel slow focus tool for the Keck I LGS-AO system can be envisioned. The expected magnitude gain is about $\Delta m = - 8.7$, although this value may not be fully realized since the ability to track the TT star could become the limiting factor. Nevertheless, the flux gain provided by GS enables slow focus tracking at higher rates. For example, with a star of $\sim$ 16 in magnitude, the LBWFS corrects slow focus every 30 s, whereas GS would allow corrections as often as every 1 s, reducing the lag error by $\sim$ 37 nm rms.

The tests performed for slow focus tracking at the Keck I telescope are an excellent case study of highly challenging conditions for the usage of FPWFS for slow focus tracking; namely, sub 0.5 Nyquist sampling, low S/N, and the presence of high-order residuals. The usage of FPWFS under these extreme conditions, which are nevertheless expected in on-sky operation, can be an important proof of concept for LGS-AO systems on existing 8-10 m telescopes and future ELTs.

\begin{acknowledgements}

Co-funded by the European Union under the Marie Skłodowska-Curie Grant Agreement No 101081465 (AUFRANDE). Views and opinions expressed are however those of the author(s) only and do not necessarily reflect those of the European Union or the Research Executive Agency. Neither the European Union nor the Research Executive Agency can be held responsible for them. \\

This work benefited from the support the French National Research Agency (ANR) with APPLY (ANR-19-CE31-0011) and LabEx FOCUS (ANR-11-LABX-0013), the Programme Investissement Avenir F-CELT (ANR-21-ESRE-0008), the Action Spécifique Haute Résolution Angulaire (ASHRA) of CNRS/INSU co-funded by CNES, the french government under the France 2030 investment plan (cassiopée project) and the Initiative d’Excellence d’Aix-Marseille Université A*MIDEX, program number AMX-22-RE-AB-151. \\

This work was performed in support of the KAPA project funded by the U.S. National Science Foundation (NSF) Mid-Scale Innovations Program award AST-1836016. The TRICK camera was developed with funding from NSF Advanced Technology and Instrumentation Program award AST-1007058. \\

Some of the data presented herein were obtained at Keck Observatory, which is a private 501(c)3 non-profit organization operated as a scientific partnership among the California Institute of Technology, the University of California, and the National Aeronautics and Space Administration. The Observatory was made possible by the generous financial support of the W. M. Keck Foundation. \\

The authors wish to recognize and acknowledge the very significant cultural role and reverence that the summit of Maunakea has always had within the Native Hawaiian community. We are most fortunate to have the opportunity to conduct observations from this mountain. \\

C.C acknowledges support funding 2022.01293.CEECIND/CP1733/CT0012 from the Portuguese Fundac\~ao para a Ci\^encia e a Tecnologia. 
    
\end{acknowledgements}

\bibliographystyle{aa} 
\bibliography{bibliography} 

\newpage
\begin{appendix}

\section{\rev{GS algorithm}}\label{Appendix_GS_algo}

\begin{figure}[h]
\centering
\includegraphics[scale=0.35]{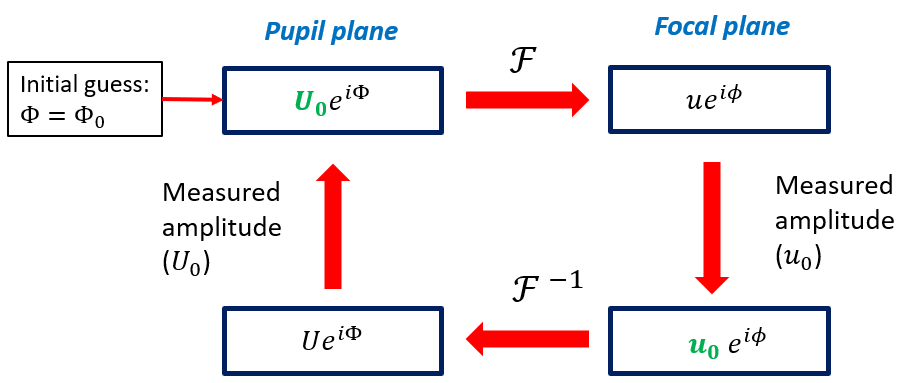}
\caption{Illustration of the GS algorithm used in this paper. This implementation corresponds to the error-reduction algorithm as defined by \cite{fienup1982}, also commonly referred to in the literature as the classic GS algorithm.}
\label{GS_algo_scheme}
\end{figure}

In the literature, several variants of the GS algorithm have been proposed. For the sake of clarity, this appendix provides a detailed description of the specific GS implementation adopted throughout this work. The version employed corresponds to the classic (or general) GS algorithm \citep{gerchberg1972}, also known as the\del{\emph{error-reduction}} \edrev{error-reduction} algorithm, as defined in Section II of \cite{fienup1982}. A schematic representation of the implemented GS algorithm is shown in Fig. \ref{GS_algo_scheme}.

\begin{figure}[h]
\centering
\includegraphics[scale=0.50]{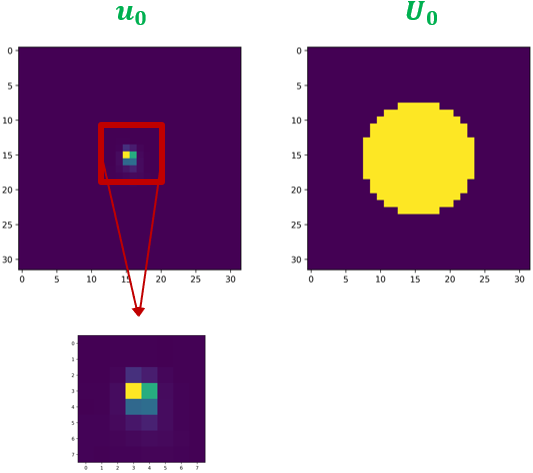}
\caption{\rev{Illustration of an example of the amplitude constraints applied in the GS algorithm at the focal plane (left) and pupil plane (right).}}
\label{GS_constraints}
\end{figure}

\rev{The algorithm consists in four simple steps performed at each iteration. These steps involve propagating the electric field back and forth between the pupil and focal planes using the Fourier transform and its inverse, while enforcing amplitude constraints in each plane. In the pupil plane, the constraint is imposed by applying the pupil mask, denoted $U_0$. In the focal plane, the amplitude constraint is given by $u_0 = \sqrt{p}$ for single point sources (see Eq. \ref{Eq_psf}). An illustrative example of these constraints for the TRICK case is shown in Fig. \ref{GS_constraints}.}

\rev{A pupil size of $16 \times 16$ pixels was found to provide the optimal trade-off between computational speed and focus-estimation accuracy. With a zero-padding factor of 2 corresponding to a Nyquist-sampled model - the pupil amplitude constraint $U_0$ is therefore represented on a $32 \times 32$ grid. The focal-plane amplitude constraint must have the same dimensions, and the square root of the TRICK image is zero-padded accordingly. Although the TRICK pixels are not Nyquist sampled, results presented throughout this paper demonstrate that the linearity of the focus estimation is very well preserved. Additional discussion on the sampling error is provided in Section \ref{Appendix_sampling_error}.}

\rev{The initial phase estimate, denoted $\Phi_0$, is obtained from an in-focus image. This image is processed using the GS algorithm to retrieve an initial phase map which, in an ideal scenario, would contain only the phase diversity (200 nm rms of astigmatism) present in TRICK's optical path. In practice, however, additional aberrations are present, most notably non-common-path aberrations (NCPAs), which must also be accounted for. As a result, the GS phase retrieved from the in-focus image includes both the known phase diversity and the NCPAs. This retrieved phase is therefore used as the initial phase estimate $\Phi_0$ in subsequent GS runs.}

This procedure, commonly referred to as diversity calibration, was investigated by Kuznetsov et al. \citep{arseniy2024} in the context of the LiFT algorithm, where it was shown to significantly improve performance. In this work, we apply the same underlying principle to the GS algorithm. For the Gf method, which estimates only the focus term, the diversity calibration reduces to a simple bias calibration: any non-zero focus measured from an in-focus image is treated as a systematic offset and subtracted from all subsequent focus measurements.

\rev{During on-sky operation, the diversity calibration is performed prior to entering closed-loop operation. This calibration can be carried out during daytime bench measurements, as was done for the on-sky tests presented in Section \ref{on_sky_section}. If required, additional diversity calibrations may be performed on-sky to compensate for slow temporal variations in the NCPAs (e.g., every $\sim10$~min). Since the system operates in closed loop and the images converge toward an in-focus state, such re-calibrations should be possible.}

\section{Sampling error}\label{Appendix_sampling_error}

Simulation results show that both LiFT and GS are capable of having excellent linear behavior in focus ramps even when a considerable sampling error is made. This allowed the usage of GS at the Keck I for slow focus tracking.

\begin{figure}[h]
\centering
\includegraphics[scale=0.50]{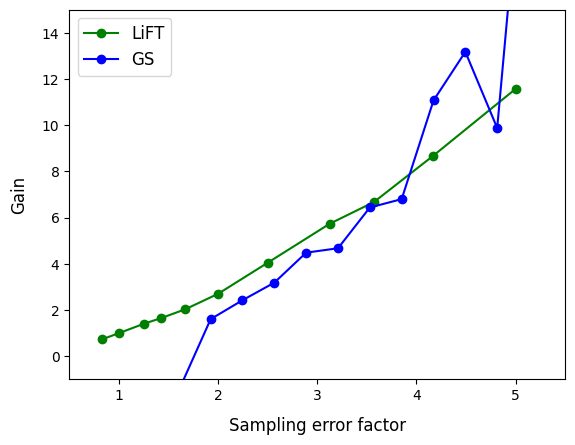}
\caption{Gain to be applied in focus ramps according to sampling error made for GS and LiFT algorithms. Results obtained in simulations using H-band.}
\label{lift_gs_gain_sampling_error}
\end{figure}

Sampling error factor is the ratio between the true angular pixel size, i.e., 50 mas, and the "model" pixel angular size. A sampling factor higher than 1 means we are overestimating the sampling. In Fig. \ref{lift_gs_gain_sampling_error}, we show the gains according to the committed sampling error. With GS, the model sampling is always at least 0.5 Nyquist, and that is why we do not cover the same sampling error range as in LiFT. As we can see, the higher the sampling error committed, the higher the gain will be, with a slope of $\sim$ 1.5-2. 

\rev{The LiFT results presented here with an intentional sampling error are included solely for academic and comparative purposes. In practice, LiFT can be applied to poorly sampled images without introducing a sampling mismatch in its model. Accordingly, all LiFT results presented in the main body of this paper are obtained using the correct sampling and do not include any form of sampling error.}

Both LiFT and GS apply nonlinear algorithms (gradient-descent like), that seek for a phase (or modes' amplitudes in LiFT case) that minimize a metric error, defined as the difference between the modeled image and the "true" image. Taking this into account, when making a sampling error, these algorithms should still be able to seek for a solution at the cost of making an error in the modes amplitudes retrieved. \rev{Low-order modes, such as focus or astigmatism, manifest themselves near the core/center of the PSF, whether the PSF is well sampled or under-sampled. This property allows the use of a model with a sampling error. In practice, the overall morphology of PSFs affected by low-order aberrations is largely preserved despite sampling inaccuracies.}

\begin{figure}[h]
\centering
\includegraphics[scale=0.35]{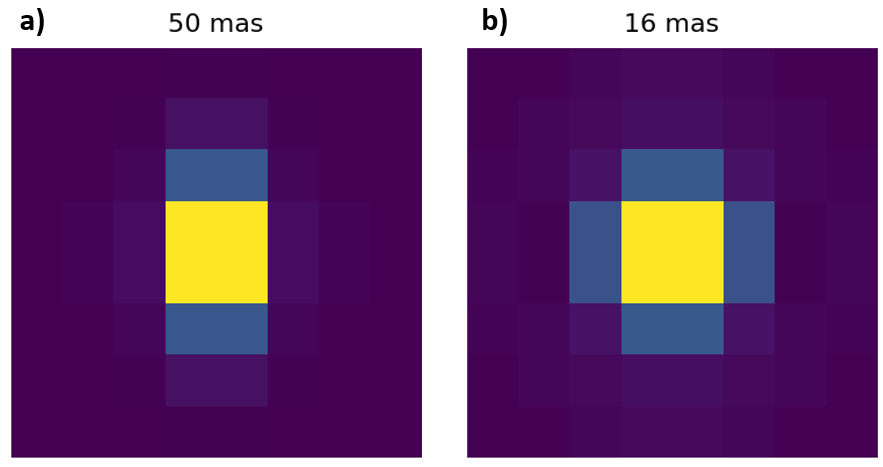}
\caption{Simulated images in H-band with astigmatism and focus: (a) 50 mas angular pixel size (TRICK case) with 200 nm rms of astigmatism and 200 nm rms of focus; (b) 16 mas angular pixel size (1 Nyquist) with 50 nm rms of astigmatism and 40 nm rms of focus.}
\label{sampling_error_visual}
\end{figure}

For example, in Fig. \ref{sampling_error_visual}, we have a visual example of two simulated images with astigmatism and focus, with angular pixel sizes of 50 mas (true sampling) and 16 mas (modeled sampling). While the image with 50 mas has 200 nm rms of both focus and astigmatism, the one with 16 mas has those same modes, but only with an amplitude of 50 nm rms of astigmatism and 40 nm rms of focus. Still, the difference between the "true" image (a) and the modeled one (b) is low, explaining why the algorithms work when making sampling errors. Despite affecting the retrieved focus values, the verified linear behavior in focus ramps, either in simulations or with bench data, was impressive, which showed the possibility of using GS for slow focus tracking at the Keck I LGS-AO system. 

\end{appendix}

\end{document}